\pdfoutput=1
\documentclass[runningheads]{llncs}
\usepackage{graphicx}
\usepackage{booktabs}
\usepackage{multirow}
\usepackage[caption=false]{subfig} 

\begin{document}
\title{\#ArsonEmergency and Australia's ``Black Summer'': Polarisation and Misinformation on Social Media}
\titlerunning{\#ArsonEmergency and Australia's ``Black Summer''}
%
\author{Derek Weber\inst{1,2}\orcidID{0000-0003-3830-9014} \and
Mehwish Nasim\inst{1,3,5,6}\orcidID{0000-0003-0683-9125} \and
Lucia Falzon\inst{4,2}\orcidID{0000-0003-3134-4351} \and
Lewis Mitchell\inst{1,5}\orcidID{0000-0001-8191-1997}}
\authorrunning{D.C. Weber et al.}
%
\institute{University of Adelaide, South Australia, Australia \\
\email{\{derek.weber,lewis.mitchell\}@adelaide.edu.au} \and
Defence Science and Technology Group, Adelaide, Australia \\
\email{derek.weber@dst.defence.gov.au} 
\and
Data61, Commonwealth Science and Industry Research Organisation, Australia \\
\email{mehwish.nasim@data61.csiro.au} 
\and
School of Psychological Sciences, University of Melbourne\\
\email{lucia.falzon@unimelb.edu.au}
\and
ARC Centre of Excellence for Mathematical and Statistical Frontiers, Australia
\and
Cyber Security Cooperative Research Centre
}
\maketitle              
\begin{abstract}






During the summer of 2019-20, while Australia suffered unprecedented bushfires across the country, false narratives regarding arson and limited backburning spread quickly on Twitter, particularly using the hashtag $\#ArsonEmergency$. 
Misinformation and bot- and troll-like behaviour were detected and reported by social media researchers 
and the news soon reached mainstream media. 
This paper examines 
the communication and behaviour of two polarised online communities before and after news of the misinformation became public knowledge.
Specifically, the \emph{Supporter} community actively engaged with others to spread the hashtag, using a variety of news sources pushing the arson narrative, while the \emph{Opposer} community engaged less, retweeted more, and focused its use of URLs to link to 
mainstream 
sources, debunking the narratives and exposing the 
anomalous 
behaviour. This influenced the content of the broader discussion. Bot analysis revealed the active accounts were predominantly human, but behavioural and content analysis suggests Supporters engaged in trolling, 
though both communities used aggressive language.

\end{abstract}
%
%
%


\section{Introduction}
\setcounter{footnote}{0} 

People share an abundance of useful information on social media during a crisis situation~\cite{BrunsLiang2012,bruns2012qldfloods}. 
This information, if analysed correctly, can rapidly reveal population-level events such as imminent civil unrest, natural disasters, or accidents~\cite{tuke2020pachinko}. 
Not all content is helpful, however: different entities may try to popularise false 
narratives using sophisticated social bots and/or humans. 
The spread of 
such misinformation not only makes it difficult for analysts to use Twitter data for public benefit~\cite{nasim2018real} but may also encourage large numbers of people to believe false narratives, which may then influence public policy and action, and can be particularly dangerous during crises~\cite{kuvsen2020you}. 

This paper presents a case study of the dynamics of misinformation propagation during one such crisis. 
The 2020 Australian `Black Summer' bushfires burnt over $16$ million hectares, destroyed over $3,500$ homes, and caused at least $33$ human and a billion animal  fatalities\footnote{https://www.abc.net.au/news/2020-02-19/australia-bushfires-how-heat-and-drought-created-a-tinderbox/11976134}, and attracted global media attention.
We show that: 

\begin{itemize}
    \item Significant Twitter discussion activity accompanied the Australian bushfires, influencing media coverage. 
    \item In the midst of this, narratives of misinformation began to circulate on social media, including that: \begin{itemize}
        \item the bushfires were caused by arson; 
        \item preventative backburning efforts were reduced due to green activism; 
        \item Australia commonly experiences such bushfires; and
        \item climate change is not related to bushfires.
    \end{itemize}
\end{itemize}

\sloppy All of these narratives were refuted, e.g., the arson figures being used were incorrect\footnote{https://www.abc.net.au/radionational/programs/breakfast/victorian-police-reject-claims-bushfires-started-by-arsonists/11857634}, preventative backburning has 
limited effectiveness
\footnote{https://www.theguardian.com/australia-news/2020/jan/08/hazard-reduction-is-not-a-panacea-for-bushfire-risk-rfs-boss-says}, 
the fires are ``unprecedented''\footnote{The Australian Academy of Science's statement: https://www.science.org.au/news-and-events/news-and-media-releases/statement-regarding-australian-bushfires}, and climate change is, in fact, increasing the frequency and severity of the fires
\footnote{Science Brief, on 14 January 2020, reports on a survey of 57 papers on the matter conducted by researchers from the University of East Anglia, Imperial College, London, Australia's CSIRO, the Univerity of Exeter and the Met Office Hadley Centre, Exeter: https://sciencebrief.org/briefs/wildfires}.
The Twitter discussion surrounding the bushfires made use of many hashtags, but according to research by Graham \& Keller~\cite{GrahamKeller2020conv} 
reported on ZDNet~\cite{Stilgherrian2020zdnet}, the arson narrative was over-represented on 
$\#ArsonEmergency$, likely created as a counter to the pre-existing 
$\#ClimateEmergency$~\cite{Barry2020mw}.
Furthermore, their research indicated that $\#ArsonEmergency$ was being boosted by bots and trolls.
This 
attracted widespread media attention, with most coverage 
debunking the arson conspiracy theory. 
This case thus presents an interesting natural experiment: the nature of the 
online narrative before the publication of the ZDnet article 
and then 
after these conspiracy theories were 
debunked.

We offer an exploratory mixed-method analysis of the Twitter activity using the term `ArsonEmergency' 
around ($\pm 7$ days) 
the publication of the ZDNet article, including comparison with 
another prominent contemporaneous bushfire-related hashtag, $\#AustraliaFire$. 
A timeline analysis revealed 
three phases of activity. 
Social network analysis of retweeting behaviour identifies two 
polarised groups of Twitter users: those promoting the arson narrative, and those exposing and arguing against it. 
These polarised groups, along with the 
unaffiliated accounts, provide a further lens through which to examine the behaviour observed. A content analysis highlights how the different groups used hashtags and 
other sources to promote their narratives. 
Finally, a brief analysis of bot-like behaviour then seeks to replicate Graham \& Keller's findings~\cite{GrahamKeller2020conv}.

Our contribution is two-fold: 1) we offer an original, focused dataset from Twitter at a critical time period covering two eras in misinformation spread\footnote{https://github.com/weberdc/socmed\_sna}; and 2) insight into the evolution of a misinformation campaign relating to the denial of climate change science and experience in dealing with bushfires.

\subsection{Related Work}

The study of 
Twitter during crises is well established~\cite{BrunsLiang2012,bruns2012qldfloods,FlewBBCS2014}, and has provided recommendations to governments and social media platforms alike 
regarding its exploitation for timely community outreach. 
The continual presence of trolling and bot behaviour diverts attention and can confuse the public at times of political significance~\cite{KellerICWSM2017,crest2017,nasim2018real,rizoiu2018debatenight} as well as creating online community-based conflict~\cite{kumar2018conflict,DattaA19conflictnetwork} and polarisation~\cite{garimella2018}.

Misinformation on social media has also been 
studied~\cite{Kumar2018FalseIO}. In particular, 
the disinformation campaign against the White Helmets rescue group in Syria is useful to consider here~\cite{StarbirdWilson2020}. 
Two clear corresponding clusters of pro- and anti-White Helmet Twitter accounts were identified 
and used 
to frame an investigation of how external references to YouTube videos and channels compared with videos embedded in Twitter. 
They found the anti-White Helmet narrative was consistently sustained through ``sincere activists'' and concerted efforts from Russian 
and alternative news sites. These particularly exploited YouTube to spread critical videos, while the pro-White Helmet activity 
relied 
on the White Helmets' own online activities and sporadic media attention. 
This interaction between supporter and detractor groups 
and the media may offer insight into activity surrounding similar crises.

\subsection{Research Questions}

Motivated by 
our observations, 
we propose the following research questions about Twitter activity during the 2019-20 Australian bushfire period:
\begin{description}
    \item [RQ1] To what extent can an online misinformation community be discerned?
    \item [RQ2] How did the spread of misinformation differ between the identified phases, and did the spread of 
    the hashtag $\#ArsonEmergency$ differ from 
    other emergent discussions (e.g., $\#AustraliaFire$)?
    \item [RQ3] How does the online behaviour of those who accept climate science differ from those who refute or question it? 
    How was it affected by media coverage exposing how the $\#ArsonEmergency$ hashtag was being used?
    \item [RQ4] To what degree was the spread of misinformation facilitated or aided by troll and/or automated bot behaviour? 
\end{description}


In the remainder of this paper, we describe our mixed-method analysis and the datasets used. A timeline analysis is followed by the polarisation analysis. The revealed polarised communities are compared from behavioural and content perspectives, as well as through bot analysis. Answers to the research questions are summarised and we conclude with observations and proposals for further study of polarised communities. 


\section{Dataset and Timeline}

The primary dataset, 
`ArsonEmergency', consists of $27,456$ tweets containing this term 
posted by $12,872$ unique accounts 
from 31 December 2019 to 17 January 2020. 
The tweets were obtained using Twitter's Standard search Application Programming Interface (API)\footnote{https://developer.twitter.com/en/docs/tweets/search/api-reference/get-search-tweets} by combining the results of searches conducted with Twarc\footnote{https://github.com/DocNow/twarc} on 8, 12, and 17 January.
As a contrast, the `AusFire' dataset comprises tweets containing the term `AustraliaFire' over the same period, made from the results of Twarc searches on 8 and 17 January. `AusFire' contains $111,966$ tweets by $96,502$ accounts. 
Broader searches using multiple related terms were not conducted due to time constraints and in the interests of comparison with Graham \& Keller's findings~\cite{GrahamKeller2020conv}. Due to the use of Twint\footnote{https://github.com/twintproject/twint} in that study, differences in dataset were possible, but expected to be minimal. 
Differences in datasets collected simultaneously with different tools have been previously noted~\cite{WeberNMF2020reliability}. 
Live filtering was also not employed, as the research started after Graham~\&~Keller's findings were reported. 

This study focuses on 
about a week of Twitter activity before and after the publication of the ZDNet article~\cite{Stilgherrian2020zdnet}. 
Prior to its publication, the narratives that arson was the primary cause of 
the bushfires and that 
fuel load 
caused the extremity of the blazes were well known in the conservative media~\cite{Barry2020mw}. 
The ZDnet article was published at 6:03am GMT (5:03pm AEST) 
on 7 January 2020, and was then reported more widely in the MSM morning news, starting around 13 hours later. 
We use these temporal markers to define three dataset phases:
\begin{itemize}
    \item \textit{Phase~1}: Before 6am GMT, 7 January 2020; 
    \item \textit{Phase~2}: From 6am to 7pm GMT, 7 January 2020; and
    \item \textit{Phase~3}: After 7pm GMT, 7 January 2020.
\end{itemize}

\begin{figure}[t!]
    \centering
    \includegraphics[width=0.8\textwidth]{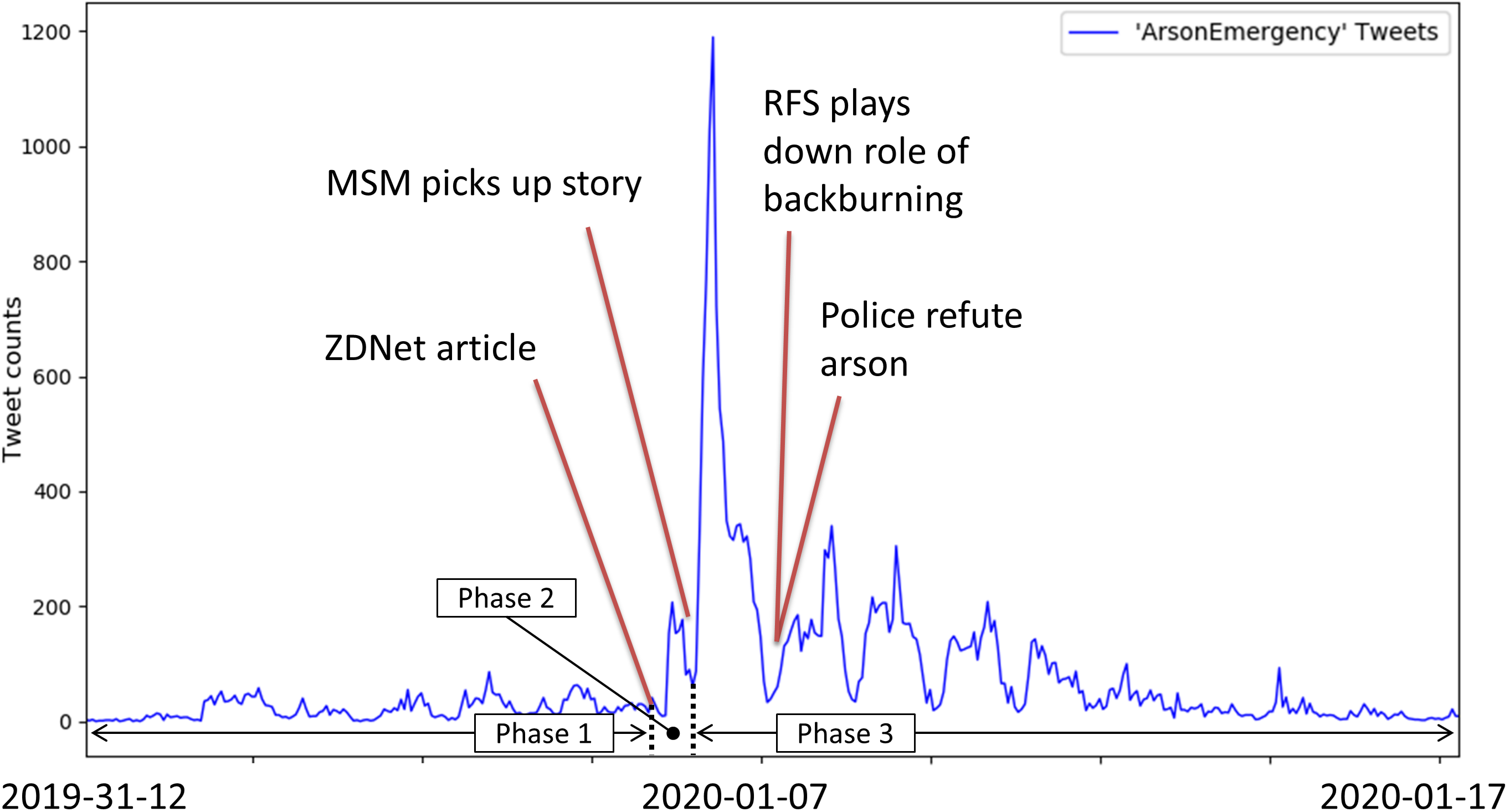}
    \caption{Tweet activity in the `ArsonEmergency' dataset, annotated with notable real-world events and the identified phases.}
    \label{fig:arson-timeline}
\end{figure}

Figure~\ref{fig:arson-timeline} shows the number of tweets posted each hour in the `ArsonEmergency' dataset, and 
highlights the 
phases and notable events including: the publication of the ZDNet article; when the story hit the MSM; the time at which the Rural Fire Service (RFS) and Victorian Police countered the narratives promoted on the $\#ArsonEmergency$ hashtag; and the clear subsequent day/night cycle. 
The RFS and Victorian Police announcements countered the false narratives promoted in political discourse in the days prior.

\sloppy Since late September 2020, Australian and international media had reported on the bushfires around Australia, including 
stories and photos drawn directly from social media, as those caught in the fires shared their experiences. 
No one hashtag had emerged to dominate the online conversation 
and many were in use, including  $\#AustraliaFires$, $\#ClimateEmergency$, $\#bushfires$, and $\#AustraliaIsBurning$. 

The use of $\#ArsonEmergency$ was limited in Phase~1, with the busiest hour having around $100$ tweets, but 
there was an influx of new accounts in Phase~2. Of all $927$ accounts active in Phase~2 (responsible for $1,207$ tweets), $824$ ($88.9\%$) of them had not posted in Phase~1 (which had $2,061$ active accounts). Content analyses revealed $1,014$ ($84\%$) of the tweets in Phase~2 were retweets, more than $60\%$ of which were retweets promoting the ZDNet article and the findings it reported. Closer examination of the timeline  
revealed that the majority of the discussion occurred between 9pm and 2am AEST, possibly inflated by a single tweet referring to the ZDNet article (at 10:19 GMT), which was retweeted $357$ times. In Phase~3, more new accounts joined the conversation, but the day/night cycle indicates that the majority of discussion was local to Australia (or at least its major timezones).

\begin{figure}[t!]
    \centering
    \includegraphics[width=0.7\textwidth]{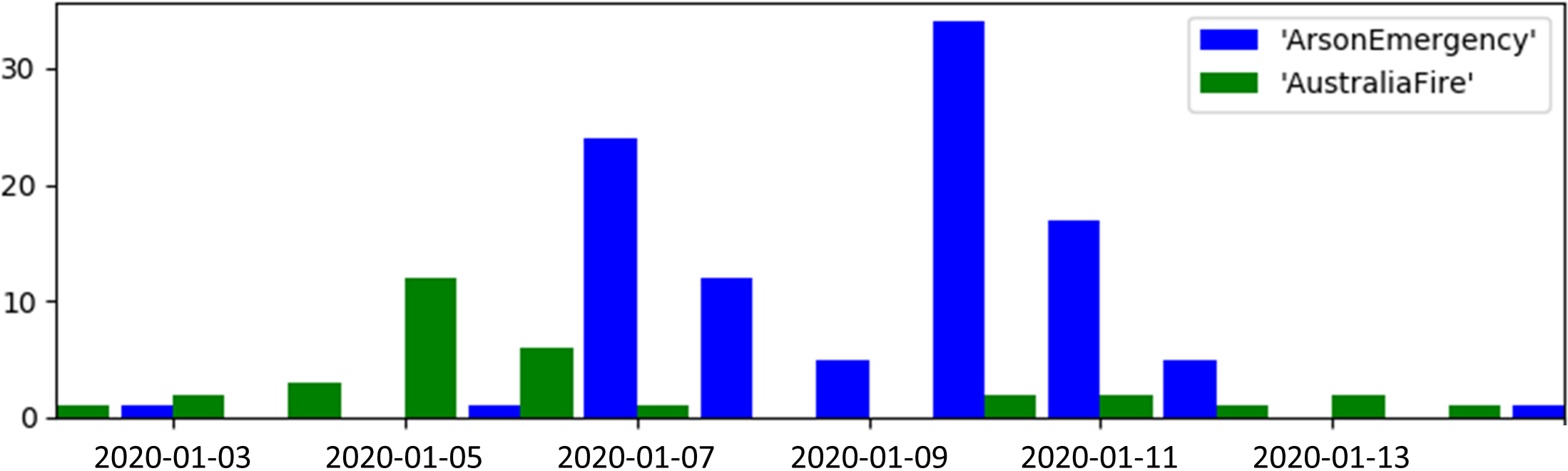}
    \caption{Counts of tweets using the terms `ArsonEmergency' and `AustraliaFire' without a `\#' symbol from 2--15 January 2020 in meta-discussion regarding each term's use as a hashtag (counts outside were zero).}
    \label{fig:arson-ausfire-term_only-counts}
\end{figure}

The term `ArsonEmergency' (sans `\#') was used for the Twarc searches, rather than `\#ArsonEmergency', to capture tweets that did not include the hashtag but were relevant to the discussion. 
Of the $27,546$ tweets in the `ArsonEmergency' dataset, only $100$ did not use it with the `\#' symbol, and only $34$ of the $111,966$ `AustraliaFire' tweets did the same. 
Figure~\ref{fig:arson-ausfire-term_only-counts} shows the
emergence of the reflexive discussion 
generated by those 
conversing about 
the discussion on $\#ArsonEmergency$ without promulgating the hashtag itself. 

\section{Polarisation in the Retweet Network} \label{sec:polarisation} 

\begin{figure}[ht!]
\centering
	\includegraphics[width=0.99\columnwidth]{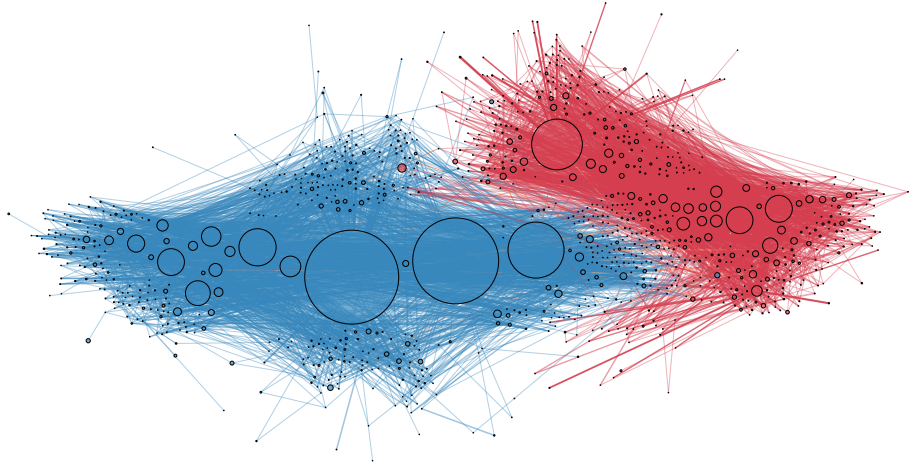}
	\caption{Polarised retweets graph about the arson theory. Left(blue): \emph{Opposers}, right(red): \emph{Supporters} of the arson narrative. Nodes represent users. An edge between two nodes means one retweeted the tweet of the other. Node size corresponds to degree centrality.}
	\label{fig:retweets-polarised}
\end{figure}

There is no agreement on whether retweets imply endorsement or alignment. Metaxas \emph{et al.}~\cite{metaxas2015retweets} studied retweeting behaviour in detail by conducting user surveys and studying over $100$ relevant papers referring to retweets. Their findings conclude that when users retweet, it indicates interest and agreement as well as trust in not only the message content but also in the originator of the tweet. This opinion is not shared by some celebrities and journalists who put a disclaimer on their profile: ``retweets $\neq$ endorsements''. Metaxas \emph{et al.}~\cite{metaxas2015retweets} also indicated that inclusion of hashtags strengthens the agreement, especially for political topics. Other motivations, such as the desire to signal to others to form bonds and manage appearances~\cite{falzon2017representation}, serve to further imply that even if retweets are not endorsements, we can assume they represent agreement or an appeal to likemindedness at the very least. 

We conducted an exploratory analysis on the retweets graph shown in Figure~\ref{fig:retweets-polarised}. The nodes indicate Twitter accounts. An edge between two accounts shows that one retweeted a tweet of the other. Using conductance cutting~\cite{BrandesGW2015clustering}, we discovered two distinct well-connected communities, with a very low number of edges between the two communities. Next, we selected the top ten accounts from each community based upon the degree centrality (most retweeted), 
manually checked their profiles, and hand labelled them as 
\emph{Supporters} and \emph{Opposers} of the arson narrative\footnote{Labelling was conducted by the first two authors independently and then compared.}. 
The accounts have been coloured accordingly in Figure~\ref{fig:retweets-polarised}: red nodes are accounts that promoted the narrative, while blue nodes are accounts that opposed them.

$\#ArsonEmergency$ had different connotations for each community. Supporters used the hashtag to reinforce their existing belief about climate change, while Opposers used this hashtag to refute the arson theory. 
The arson theory was a topic on which people held strong opinions resulting in the formation of the two strongly connected communities. Such polarised communities typically do not admit much information flow between them, hence members of such communities are repeatedly exposed to similar narratives, which further strengthens their existing beliefs. Such closed communities are also known as \emph{echo chambers}, and they limit people's information space. The retweets tend to coalesce within communities, as has been shown for Facebook comments~\cite{nasim2013commenting}. 

These two groups, Supporters and Opposers, and those users unaffiliated with either group, are used to frame the remainder of the analysis in this paper.




\subsection{Behaviour}

User behaviour on Twitter can be examined through the features used to connect with others and through content. Here we consider how active the different groups were across the phases of the collection, and then how that activity manifested itself in the use of mentions, hashtags, URLs, replies, quotes and retweets. 

\begin{table}
    \centering\small
    \caption{Activity of the polarised retweeting accounts, by interaction type broken down by phase.}
    \label{tab:group-activity-by-interaction-type-and-phase}
    \resizebox{\columnwidth}{!}{%
        \begin{tabular}{@{}p{1.5em}|llrrrrrrrr@{}}
            \toprule
            & Group &                    & Tweets & Accounts & Hashtags & Mentions & Quotes & Replies & Retweets & URLs  \\
            \midrule
            \multirow{4}{*}{\rotatebox[origin=c]{90}{\textbf{Phase~1}}} 
            & \multirow{2}{*}{Supporters} 
                    & \emph{Raw count}   &  1,573 &      360 &    2,257 &    2,621 &    185 &     356 &      938 &   405 \\
            &       & \emph{Per account} &  4.369 &      --- &    1.435 &    1.666 &  0.118 &   0.226 &    0.596 & 0.257 \\
                    \cmidrule{3-11}
            & \multirow{2}{*}{Opposers} 
                    & \emph{Raw count}   &     33 &       21 &      100 &       35 &      8 &       2 &       20 &     9 \\
            &       & \emph{Per account} &  1.571 &      --- &    3.030 &    1.061 &  0.242 &   0.061 &    0.606 & 0.273 \\
            \midrule
            \multirow{4}{*}{\rotatebox[origin=c]{90}{\textbf{Phase~2}}} 
            & \multirow{2}{*}{Supporters} 
                    & \emph{Raw count}   &    121 &       77 &      226 &      159 &     11 &      29 &       74 &    24 \\
            &       & \emph{Per account} &  1.571 &      --- &    1.868 &    1.314 &  0.091 &   0.240 &    0.612 & 0.198 \\
                    \cmidrule{3-11}
            & \multirow{2}{*}{Opposers} 
                    & \emph{Raw count}   &    327 &      172 &      266 &      476 &      7 &      14 &      288 &    31 \\
            &       & \emph{Per account} &  1.901 &      --- &    0.813 &    1.456 &  0.021 &   0.043 &    0.881 & 0.095 \\
            \midrule
            \multirow{4}{*}{\rotatebox[origin=c]{90}{\textbf{Phase~3}}} 
            & \multirow{2}{*}{Supporters} 
                    & \emph{Raw count}   &  5,278 &      474 &    7,414 &    7,407 &    593 &   1,159 &    3,212 &   936 \\
            &       & \emph{Per account} & 11.135 &      --- &    1.405 &    1.403 &  0.112 &   0.220 &    0.609 & 0.177 \\
                    \cmidrule{3-11}
            & \multirow{2}{*}{Opposers} 
                    & \emph{Raw count}   &  3,227 &      585 &    3,997 &    3,617 &    124 &      95 &    2,876 &   359 \\
            &       & \emph{Per account} &  5.516 &      --- &    1.239 &    1.121 &  0.038 &   0.029 &    0.891 & 0.111 \\
            \midrule
            \multirow{4}{*}{\rotatebox[origin=c]{90}{\textbf{Overall}}} 
            & \multirow{2}{*}{Supporters} 
                    & \emph{Raw count}   &  6,972 &      497 &    9,897 &   10,187 &      789 &  1,544 &   4,224 &  1,365 \\
            &       & \emph{Per account} & 14.028 &      --- &    1.420 &    1.461 &    0.113 &  0.221 &   0.606 &  0.196 \\
                    \cmidrule{3-11}
            & \multirow{2}{*}{Opposers} 
                    & \emph{Raw count}   &  3,587 &      593 &    4,363 &    4,128 &      139 &    111 &   3,184 &   399 \\
            &       & \emph{Per account} &  6.049 &      --- &    1.216 &    1.151 &    0.039 &  0.031 &   0.888 & 0.111 \\
            \bottomrule
        \end{tabular}
    }
\end{table}

Considering each phase (Table~\ref{tab:group-activity-by-interaction-type-and-phase}) Supporters used $\#ArsonEmergency$ nearly fifty times more often than Opposers, which accords with Graham \& Keller's findings that the false narratives were significantly more prevalent on that hashtag compared with others in use at the time~\cite{Stilgherrian2020zdnet,GrahamKeller2020conv}. In Phase~2, during the Australian night, Opposers countered with three times as many tweets as Supporters, including fewer hashtags, more retweets, and half the number of replies, demonstrating different behaviour to Supporters, which actively used the hashtag in conversations. 
Content analysis confirmed this to be the case. This is evidence that Supporters wanted to promote the hashtag to promote the narrative. Interestingly, Supporters, having been relatively quiet in Phase~2, produced 64\% more tweets in Phase~3 than Opposers, 
using proportionately more of all interactions except retweeting, and many more replies, quotes, and tweets spreading the narrative by using multiple hashtags, URLs and mentions. In short, Opposers tended to rely more on retweets, while Supporters engaged directly 
and were more active in the longer phases.

The concentration of narrative from certain voices requires attention. 
To consider this, Table~\ref{tab:retweet-concentration-by-phase-and-group} shows the degree to which accounts were retweeted by the different groups by phase and overall. 
Unaffiliated accounts relied on a smaller pool of accounts to retweet than both Supporters and Opposers in each phase and overall, which is reasonable to expect as the majority of Unaffiliated activity occurred in Phase~3, once the story reached the mainstream news, and therefore had access to tweets about the story from the media and prominent commentators. 
Of the top $41$ retweeted accounts, which retweeted $100$ times or more in the dataset, $17$ were Supporters and $20$ Opposers. Supporters were retweeted $5,487$ times ($322.8$ retweets per account), while Opposers were retweeted $8,833$ times ($441.7$ times per account). Together, affiliated accounts contributed $93.3\%$ of the top $41$'s $15,350$ retweets, in a dataset with $21,526$ retweets overall, and the top $41$ accounts were retweeted far more often than most. 
This pattern was also apparent in the $25$ accounts most retweeted by Unaffiliated accounts in Phase~3 (accounts retweeted at least $100$ times): $8$ were Supporters and $14$ were Opposers.
Thus Supporters and Opposers made up the majority of the most retweeted accounts, and arguably influenced the discussion more than Unaffiliated accounts.

\begin{table}
    \centering\small
    \caption{Retweeting activity in the dataset, by phase and group.}
    \label{tab:retweet-concentration-by-phase-and-group}
    \resizebox{\columnwidth}{!}{%
        \begin{tabular}{@{}c|rrr|rrr|rrr@{}}
            \toprule
                    & \multicolumn{3}{c}{Supporters}      & \multicolumn{3}{c}{Opposers}        & \multicolumn{3}{c}{Unaffiliated}    \\
            Phase   & Retweets & Retweeted & Retweets per & Retweets & Retweeted & Retweets per & Retweets & Retweeted & Retweets per \\
                    &          & Accounts  & account      &          & Accounts  & account      &          & Accounts  & account      \\
            \cmidrule(r){1-1} \cmidrule(lr){2-4} \cmidrule(l){5-7} \cmidrule(l){8-10} 
            1       &      938 &        77 &       12.182 &       20 &         8 &        2.500 &    1,659 &       105 &       15.800 \\
            2       &       74 &        21 &        3.524 &      288 &        31 &        9.290 &      652 &        60 &       10.867 \\
            3       &    3,212 &       290 &       11.076 &    2,876 &       228 &       12.614 &   11,807 &       532 &       22.194 \\ 
            \midrule
            Overall &    4,224 &       327 &       12.917 &    3,184 &       243 &       13.103 &   14,118 &       613 &       23.030 \\
            \bottomrule        
        \end{tabular}
    }
\end{table}




\subsection{Content}

When contrasting the content of the two affiliated groups, we considered the hashtags and external URLs used. 
A hashtag can provide a proxy for a tweet's
topic, and an external URL can refer a tweet's reader to further 
information relevant to the tweet, and therefore tweets that use the same URLs and
hashtags can be considered related.



\subsubsection{Hashtags.}

\begin{figure}[!ht]
    \subfloat[Supporter hashtags.\label{fig:ccd_hashtag_co-mentions}]{%
        \includegraphics[width=0.4557\textwidth]{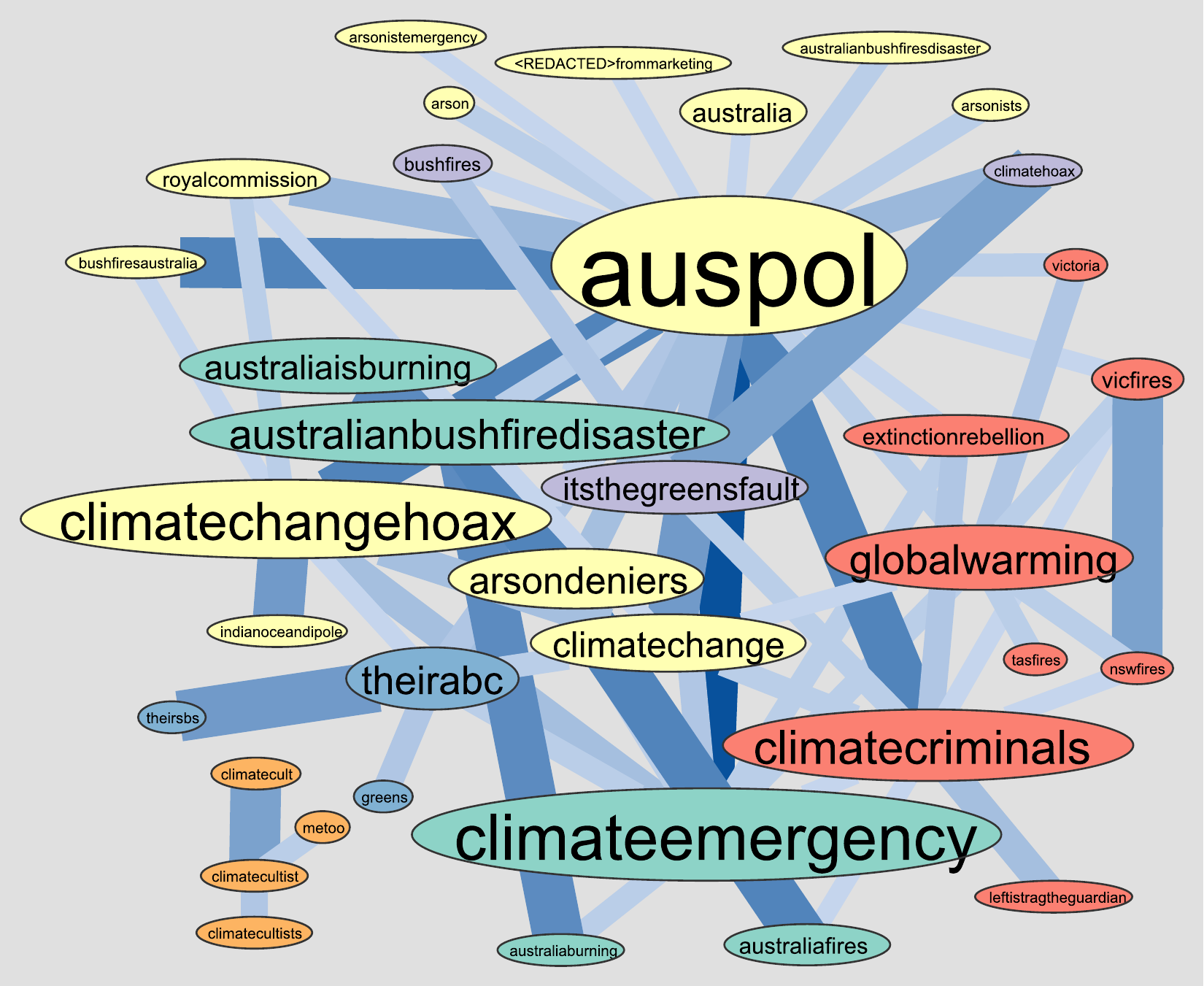}
    }
    \hfill
    \subfloat[Opposer hashtags.\label{fig:cca_hashtag_co-mentions}]{%
        \includegraphics[width=0.525\textwidth]{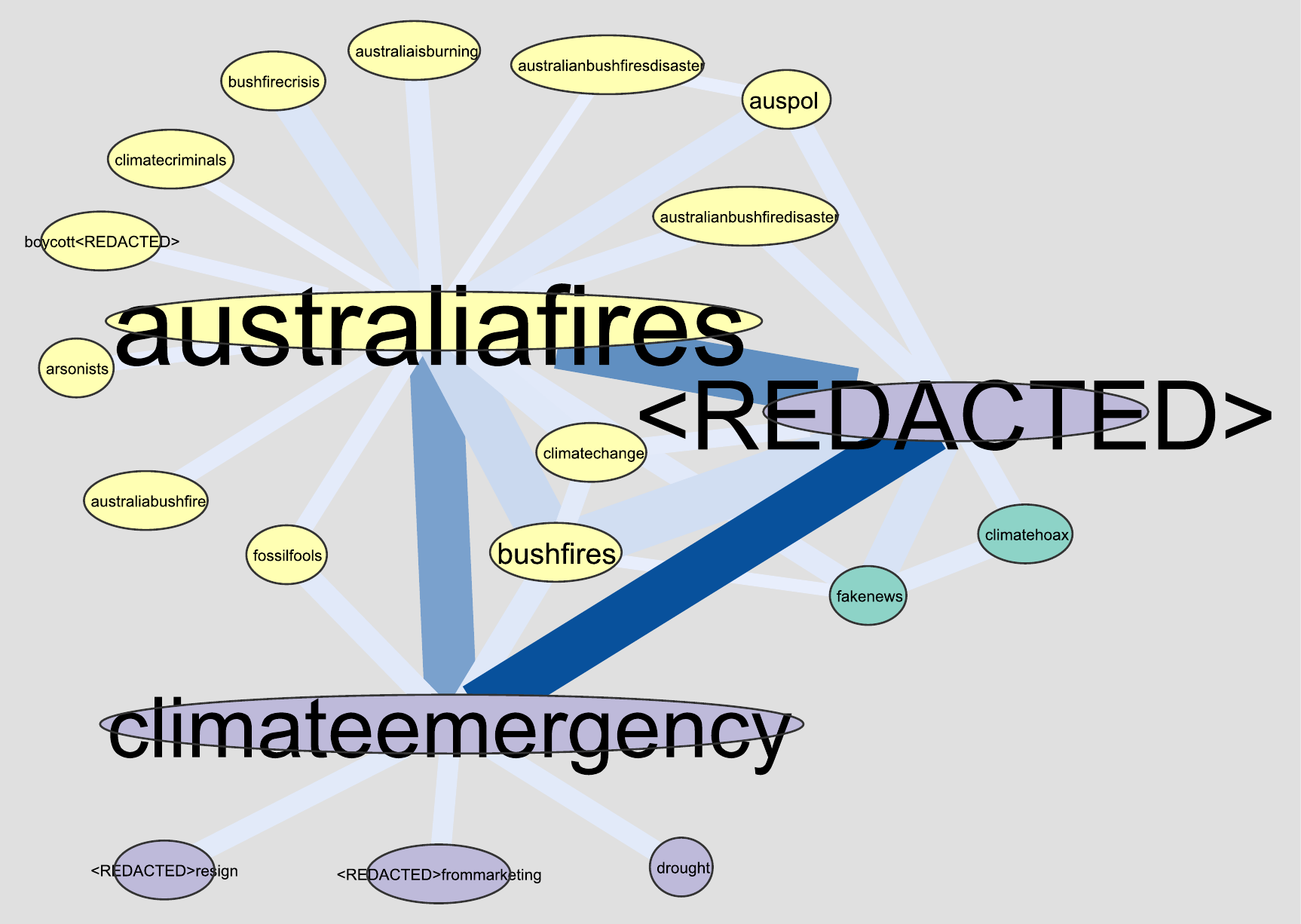}
    }
    \caption{Co-mentioned hashtags of Supporters and Opposers. Hashtag nodes are linked when five or more accounts tweeted both hashtags, and are coloured by cluster. \textless REDACTED\textgreater~hashtags include identifying information. Heavy edges (with high weight) are thicker and darker.}
    \label{fig:hashtag_co-mentions}
\end{figure}

To discover \emph{how} hashtags were used, rather than simply \emph{which} were used, we developed co-mention graphs (Figure~\ref{fig:hashtag_co-mentions}). 
Each node is a hashtag, sized by degree centrality; 
edges represent 
an account using both hashtags (not necessarily in the same tweet); the edge weight represents the number of such accounts in the dataset. 
Nodes are coloured according to cluster detected with the widely used Louvain method~\cite{blondel2008}. 
We removed the $\#ArsonEmergency$ hashtag (as nearly each tweet in the dataset contained it) as well as edges having weight less than $5$. 
Opposers used a smaller set of hashtags, predominantly linking $\#AustraliaFires$ with $\#ClimateEmergency$ and a hashtag referring to a well-known publisher. 
In contrast, Supporters used a variety of hashtags in a variety of combinations, mostly focusing on terms related to `fire', but only a few with `arson' or `hoax', and linking to $\#auspol$ and $\#ClimateEmergency$.
Manual review of Supporter tweets included many containing only a string of hashtags, unlike the Opposer tweets. Notably, the $\#ClimateChangeHoax$ node 
has a similar degree to the $\#ClimanteChangeEmergency$ node, 
indicating Supporters' skepticism of the science, but perhaps also attempts by Supporters to join or merge the communities.

Manual inspection of Supporter tweets revealed that replies often consisted solely of ``$\#ArsonEmergency$'' (one Supporter replied to an Opposer $26$ times in under $9$ minutes with a tweet just consisting of the hashtag, although in six of the tweets @mentions of other influential Twitter accounts were also included). 
This kind of behaviour, in addition to inflammatory language in other Supporter replies, suggests a degree of aggression, though aggressive language was also noted among Opposers. Only $1.7\%$ of Opposer tweets included more than $5$ hashtags, while $2.8\%$ of Supporter
ones did, compared with $2.1\%$ 
unaffiliated.

\subsubsection{External URLs.}

URLs 
in tweets can be categorised as \emph{internal} or \emph{external}. 
Internal URLs refer to other tweets in retweets or quotes
, while external URLs are often included to highlight something about their content, 
e.g., as a source to support a claim. By analysing the URLs, 
it is possible to gauge the intent of the tweet's author by considering the reputation of the source or the 
argument offered. 

We categorised\footnote{Categorisation was conducted by two authors and confirmed by the others.} 
the top ten URLs used most by Supporters, Opposers, and the unaffiliated across the three phases, 
and found a 
significant difference between the groups. URLs were categorised into four categories:
\begin{description}
    \item[NARRATIVE] 
    Articles used to emphasise the conspiracy narratives by prominently reporting arson figures and fuel load discussions.
    \item[CONSPIRACY] Articles and web sites that 
    take extreme positions on climate change (typically arguing against predominant scientific opinion).
    \item[DEBUNKING] News articles providing authoritative information about the bushfires and related misinformation on social media.
    \item[OTHER] Other web pages.
\end{description}

\begin{figure}[!ht]
     \subfloat[Opposer URLs.\label{fig:cca_urls_by_phase}]{%
       \includegraphics[width=0.32\textwidth]{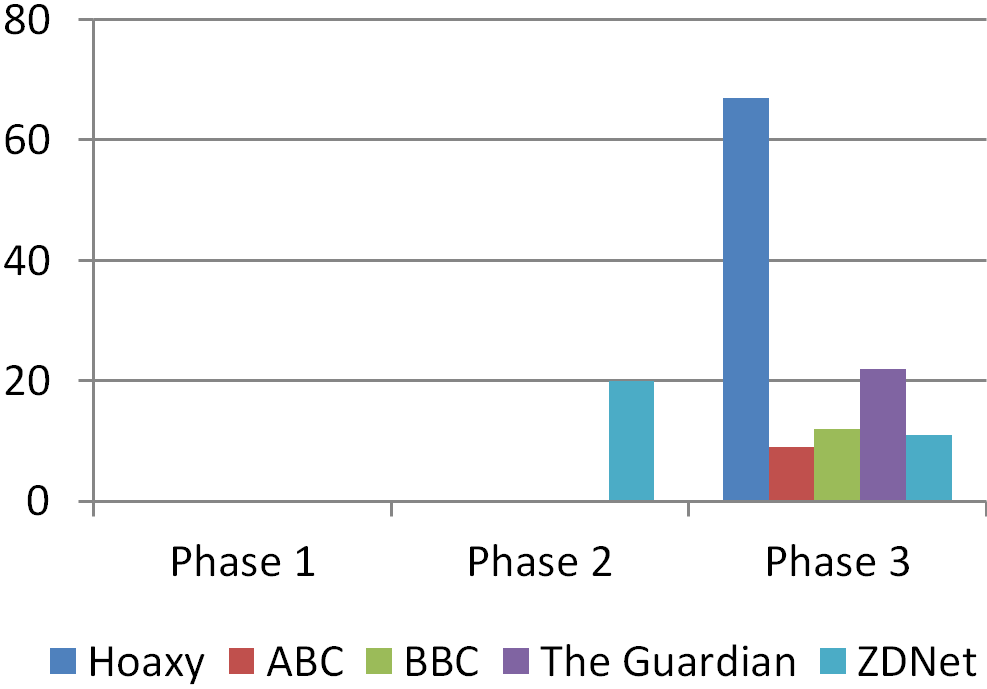}
     }
     \hfill
     \subfloat[Supporter URLs.\label{fig:ccd_urls_by_phase}]{%
       \includegraphics[width=0.32\textwidth]{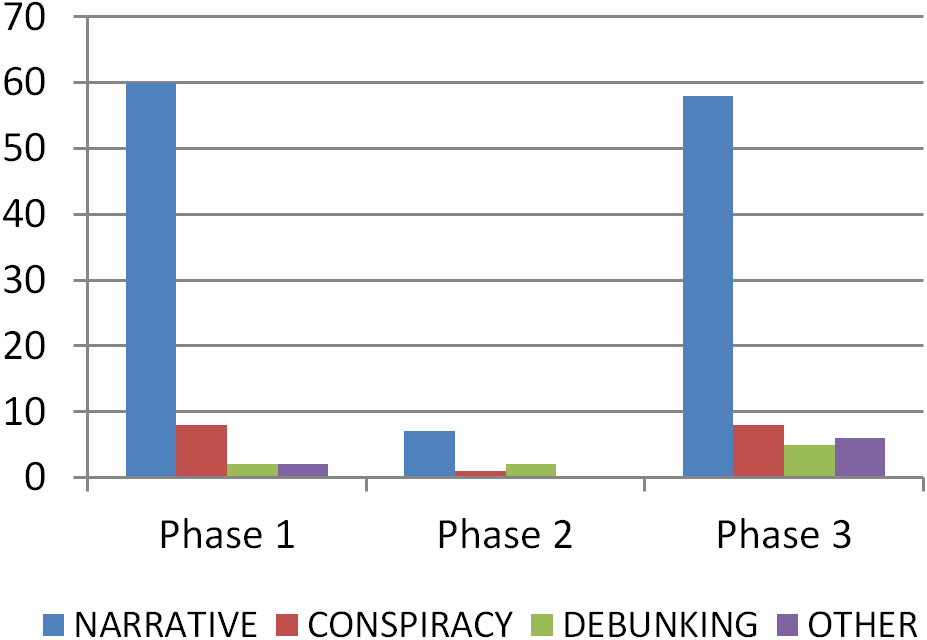}
     }
     \hfill
     \subfloat[Unaffiliated URLs.\label{fig:unaff_urls_by_phase}]{%
       \includegraphics[width=0.32\textwidth]{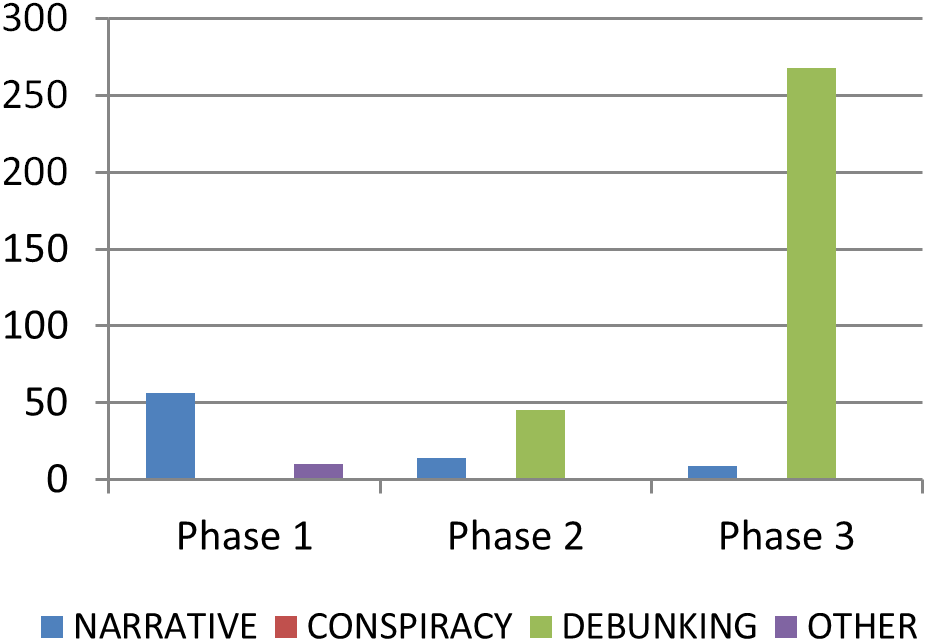}
     }
     \caption{URLs used by Opposers, Supporters and unaffiliated accounts.}
     \label{fig:urls_by_phase}
\end{figure}

URLs posted by Opposers were concentrated in Phase~3 and were all in the DEBUNKING category, with nearly half attributed to Indiana University's Hoaxy service~\cite{Shao_2016}, 
and nearly a quarter referring to the original ZDNet article~\cite{Stilgherrian2020zdnet} 
(Figure~\ref{fig:cca_urls_by_phase}). In contrast, Supporters used many URLs in Phases~1 and~3, focusing mostly on articles 
emphasising the arson narrative, 
but with references to a number of climate change denial or right wing blogs and news sites (Figure~\ref{fig:ccd_urls_by_phase}).

Figure~\ref{fig:unaff_urls_by_phase} shows that the media coverage changed the content of the unaffiliated discussion, 
from articles 
emphasising the arson
narratives in Phase~1 to 
Opposer-aligned articles 
in Phase~3. Although the activity of Supporters in Phase~3 increased significantly, the unaffiliated members appeared to refer to Opposer-aligned external URLs much more often.

Supporters used many more URLs than Opposers overall ($1,365$ to $399$) and nearly twice as many external URLs ($390$ to $212$). Supporters seemed to use many different URLs in Phase~3 and overall, but focused much more on particular URLs in Phase~1. 
Of the total number of unique URLs used in Phase~3 and overall, $263$ and $390$, respectively, only $77$ ($29.3\%$) and $132$ ($33.8\%$) appeared in the top ten, implying a wide variety of URLs were used. In contrast, in Phase~1, $72$ of $117$ appeared in the top ten ($61.5\%$), similar to Opposers' $141$ of $212$ ($66.5\%$), implying a greater focus on specific sources of information. In brief, it appears Opposers overall and Supporters in Phase~1 were 
focused in their choice of sources, but by Phase~3, Supporters had expanded their range 
considerably.



\section{Botness Analysis}

The analysis reported in ZDNet~\cite{Stilgherrian2020zdnet} indicated 
widespread bot-like behaviour 
by using \texttt{tweetbotornot}\footnote{https://github.com/mkearney/tweetbotornot}. 
Our re-analysis of this finding had two goals: 1)~attempt to replicate Graham~\&~Keller's findings in Phase 1 of our dataset; and 2)~examine the contribution of bot-like accounts detected in Phase~1 in the other phases. 
Specifically, we considered the questions:
\begin{itemize}
    \item Does another bot detection system find similar levels of bot-like behaviour?
    \item Does the behaviour of any bots from Phase~1 change in Phases~2 and~3?
\end{itemize}

We evaluated $2,512$ or $19.5\%$ of the accounts in the dataset using Botometer~\cite{botornot2016}, including all Supporter and Opposer accounts, plus all accounts that posted at least three tweets either side of Graham~\&~Keller's analysis 
reaching 
the MSM.





Botometer~\cite{botornot2016} is an ensemble bot classifier for Twitter accounts, relying on over a thousand features drawn from six categories. 
It includes a ``Complete Automation Probability'' (CAP), a Bayesian-informed probability that the account in question is ``fully automated''. 
This does not accommodate hybrid accounts~\cite{GrimmeAA2018changingperspectives} and only uses English training data~\cite{nasim2018real}, leading some researchers to use conservative ranges of CAP scores for high confidence that an account is human (\textless $0.2$) or bot (\textgreater $0.6$)~\cite{rizoiu2018debatenight}. We adopt that categorisation.

\begin{table}
    \centering\small
    \caption{Botness scores and contribution to the discussion across the phases.}
    \label{tab:botness-scores-and-contribution-by-phase}
        \begin{tabular}{@{}l|c|r|rrr|rrr@{}}
            \toprule
                      &           &       & \multicolumn{3}{c}{Active accounts} & \multicolumn{3}{c}{Tweets contributed} \\ 
            Category  & CAP       & Total & Phase~1 & Phase~2 & Phase~3         & Phase~1 & Phase~2 & Phase~3 \\
            \midrule
            Human	  & $0.0-0.2$ & 2,426 &	    898 &     438 &           1,931 &   2,213 &     674 &  11,700 \\
            Undecided &	$0.2-0.6$ &    66 &      20 &       6 &              56 &      28 &      11 &     304 \\
            Bot	      & $0.6-1.0$ &    20 &       9 &       4 &              11 &      23 &       6 &      84 \\
            \bottomrule
        \end{tabular}
\end{table}

Table~\ref{tab:botness-scores-and-contribution-by-phase} shows that the majority of accounts were human and contributed more than any automated or potentially automated accounts. 
This contrast with the reported findings~\cite{Stilgherrian2020zdnet} may be due to a number of reasons. 
The CAP score is focused on non-hybrid, English accounts, whereas \texttt{tweetbotornot} may provide a more general score, taking into account troll-like behaviour. The content and behaviour analysis discussed above certainly indicates Supporters engaged more with replies and quotes, consistent with other observed trolling behaviour~\cite{kumar2018conflict} or ``sincere activists''~\cite{StarbirdWilson2020}. 
The collection tool used, Twint, may have obtained different tweets to Twarc, as it explicitly avoids Twitter's APIs. It is possible its avoidance of the API reveals more bot-like behaviour. 
Finally, it is unclear what Graham~\&~Keller's collection strategy was; if it focused on the 
particular 
accounts which drew their attention to $\#ArsonEmergency$ to begin with, it may not have included the wider range of behaviour evident in our dataset.

\section{Discussion}

We are now well-placed to address our research questions:
\begin{description}
    \item [RQ1] \emph{Discerning a misinformation-sharing community.} 
    Analysis 
    revealed two distinct polarised communities. The content posted by the most influential accounts in these communities shows Supporters were responsible for the majority of arson-related content, 
    while Opposers 
    countered the arson narrative. 

    \item [RQ2] \emph{Differences in the spread of misinformation across phases and other discussions.} Considering URL and hashtag use in Phase~1 and~3, while the number of active Supporters grew from $360$ to $474$, the number of unique external URLs they used more than doubled, from $117$ to $263$. This was possibly due to the increased traffic on 
    $\#ArsonEmergency$.
    The number of hashtags 
    increased from $182$ hashtags used $2,257$ times to $505$ hashtags used $7,414$ times. This implies Supporters attempted to connect $\#ArsonEmergency$ with other hashtag-based communities. In contrast, Opposer activity increased from $33$ hashtags used $100$ times to $182$ hashtags used $3,997$ times, but Figure~\ref{fig:cca_hashtag_co-mentions} shows Opposers focused the majority of their discussion on a comparatively small number of hashtags. 
    
    
    \item [RQ3] \emph{Behavioural differences over time and the impact of media coverage.} Supporters were more active in Phase~1 and~3 and used 
    more types of interaction than Opposers, 
    especially 
    replies and quotes, implying a significant degree of engagement, whether as trolls 
    or as ``sincere activists''~\cite{StarbirdWilson2020}. 
    Opposers and Supporters made up the majority of retweeted accounts overall, and made up $22$ of the top $25$ accounts retweeted by unaffiliated accounts in Phase~3. 
    Supporters' use of interaction types remained steady from Phase~1 to~3. 
    While behaviour remained relatively similar, activity grew for both groups after the story reached the MSM. The vast majority of accounts shared 
    articles debunking the false narratives. The ZDNet article also affected activity, spurring Opposers and others to share the analysis it reported.
    
    \item [RQ4] \emph{Support from bots and trolls.} We found very few bots, 
    but aggressive troll-like behaviour was observed in the Supporter community. 
    Aggressive language was observed in both affiliated groups. 
    Distinguishing deliberate baiting from honest enthusiasm (even with swearing), however, is non-trivial~\cite{StarbirdWilson2020}.
\end{description}

The $\#ArsonEmergency$ activity on Twitter in early 2020 provides a unique microcosm to study the growth of a misinformation campaign before and after it was widely known. 
Our study reveals the following:
\begin{itemize}
    \item Two clear polarised communities with distinct behaviour patterns and use of content 
    were present.

    \item Supporters were more active and more engaged. 
    Opposers relied on retweeting more, and focused on a few prominent hashtags, while Supporters used many. 
    This was possibly to widely promote their message, 
    or due to non-Australian contributors being unfamiliar with which hashtags to use for an Australian audience. 
    
    \item The majority of Phase~1 $\#ArsonEmergency$ discussion referred to articles relevant to the arson narratives, but after the story reached the MSM, only the Supporter community continued to use such links.
    
    \item The majority of unaffiliated accounts shifted focus from CCD narrative-related articles in Phase~1 to debunking sites and articles in Phase~3. It is unclear whether the change in behaviour was driven by accounts changing opinion or the influx of new accounts.
    
    \item The $\#ArsonEmergency$ growth rate followed a pattern similar to another related hashtag that appeared shortly before it ($\#AustraliaFire$).
    
    \item The influence of bot accounts appears limited when analysed with Botometer~\cite{botornot2016}. It classified $0.8\%$ ($20$ of $2,512$) of accounts as bots, and $96.6\%$ ($2,426$ of $2,512$) of the remaining accounts confidently as human. Graham~\&~Keller had found an even spread of bot scores, with an average score over $0.5$. Only $20\%$ of accounts had a score $\leq 0.2$ and $46\% \geq 0.6$~\cite{Stilgherrian2020zdnet}.
\end{itemize}

Further research is required to examine social and interaction structures formed by groups involved in spreading misinformation to learn more about how such groups operate and better address the challenge they pose to society.
Future work will draw more on social network analysis based on interaction patterns and content~\cite{bagrow2019information} as well as developing a richer, more nuanced understanding of the Supporter community itself, including more content and behaviour analysis.

\section*{Acknowledgment}
The work has been partially supported by the Cyber Security Research Centre Limited whose activities are partially funded by the Australian Government’s Cooperative Research Centres Programme.

\section*{Ethics}
All data was collected, stored and analysed in accordance with Protocols \mbox{H-2018-045} and $\#170316$ as approved by the University of Adelaide's human research ethics committee.

\bibliographystyle{abbrv}
\bibliography{ref}

\begin{thebibliography}{10}

\bibitem{bagrow2019information}
J.~P. Bagrow, X.~Liu, and L.~Mitchell.
\newblock Information flow reveals prediction limits in online social activity.
\newblock {\em Nature Human Behaviour}, 3(2):122--128, 2019.

\bibitem{Barry2020mw}
P.~Barry.
\newblock {Broadcast 3rd February 2020: News Corps Fire Fight}.
\newblock {\em Media Watch, Australian Broadcasting Corporation}, 2020(1), Feb.
  2020.

\bibitem{blondel2008}
V.~D. Blondel, J.-L. Guillaume, R.~Lambiotte, and E.~Lefebvre.
\newblock Fast unfolding of communities in large networks.
\newblock {\em Journal of Statistical Mechanics: Theory and Experiment},
  2008(10):P10008, 2008.

\bibitem{BrandesGW2015clustering}
U.~Brandes, M.~Gaertler, and D.~Wagner.
\newblock Engineering graph clustering: Models and experimental evaluation.
\newblock {\em {ACM} Journal of Experimental Algorithmics}, 12:1.1:1--1.1:26,
  2007.

\bibitem{bruns2012qldfloods}
A.~Bruns and J.~Burgess.
\newblock {\#qldfloods and @QPSMedia: Crisis Communication on Twitter in the
  2011 South East Queensland Floods}.
\newblock Research Report 48241, ARC Centre of Excellence for Creative
  Industries and Innovation, Jan. 2012.

\bibitem{BrunsLiang2012}
A.~Bruns and Y.~E. Liang.
\newblock Tools and methods for capturing {T}witter data during natural
  disasters.
\newblock {\em First Monday}, 17(4), Apr. 2012.

\bibitem{crest2017}
CREST.
\newblock {R}ussian interference and influence measures following the 2017 {UK}
  terrorist attacks.
\newblock Policy Brief 17-81-2, Centre for Research and Evidence on Security
  Threats, Cardiff University, Dec. 2017.

\bibitem{DattaA19conflictnetwork}
S.~Datta and E.~Adar.
\newblock Extracting inter-community conflicts in {R}eddit.
\newblock In {\em {ICWSM}}, pages 146--157. {AAAI} Press, 2019.

\bibitem{botornot2016}
C.~A. Davis, O.~Varol, E.~Ferrara, A.~Flammini, and F.~Menczer.
\newblock {BotOrNot}: {A} system to evaluate social bots.
\newblock In {\em {WWW} (Companion Volume)}, pages 273--274. {ACM}, 2016.

\bibitem{falzon2017representation}
L.~Falzon, C.~McCurrie, and J.~Dunn.
\newblock Representation and analysis of {T}witter activity: {A} dynamic
  network perspective.
\newblock In {\em {ASONAM}}, pages 1183--1190. {ACM}, 2017.

\bibitem{FlewBBCS2014}
T.~Flew, A.~Bruns, J.~Burgess, K.~Crawford, and F.~Shaw.
\newblock Social media and its impact on crisis communication: Case studies of
  {T}witter use in emergency management in {A}ustralia and {New Zealand}.
\newblock In {\em 2013 ICA Shanghai Regional Conference: Communication and
  Social Transformation}, Nov. 2014.

\bibitem{garimella2018}
V.~R.~K. Garimella, G.~D.~F. Morales, A.~Gionis, and M.~Mathioudakis.
\newblock Polarization on social media.
\newblock In {\em {WWW} ({Tutorial Volume})}. {ACM}, 2018.

\bibitem{GrahamKeller2020conv}
T.~Graham and T.~R. Keller.
\newblock {Bushfires, bots and arson claims: Australia flung in the global
  disinformation spotlight}.
\newblock
  https://theconversation.com/bushfires-bots-and-arson-claims-australia-flung-in-the-global-disinformation-spotlight-129556,
  Jan. 2020.
\newblock (Accessed on 2020-02-07).

\bibitem{GrimmeAA2018changingperspectives}
C.~Grimme, D.~Assenmacher, and L.~Adam.
\newblock Changing perspectives: Is it sufficient to detect social bots?
\newblock In {\em {HCI} {(13)}}, volume 10913 of {\em Lecture Notes in Computer
  Science}, pages 445--461. Springer, 2018.

\bibitem{KellerICWSM2017}
F.~B. Keller, D.~Schoch, S.~Stier, and J.~Yang.
\newblock How to manipulate social media: Analyzing political astroturfing
  using ground truth data from {S}outh {K}orea.
\newblock In {\em {ICWSM}}, pages 564--567. {AAAI} Press, 2017.

\bibitem{kumar2018conflict}
S.~Kumar, W.~L. Hamilton, J.~Leskovec, and D.~Jurafsky.
\newblock Community interaction and conflict on the web.
\newblock In {\em {WWW}}, pages 933--943. {ACM}, 2018.

\bibitem{Kumar2018FalseIO}
S.~Kumar and N.~Shah.
\newblock False information on web and social media: A survey.
\newblock {\em CoRR}, abs/1804.08559, 2018.

\bibitem{kuvsen2020you}
E.~Ku{\v{s}}en and M.~Strembeck.
\newblock You talkin’ to me? {E}xploring human/bot communication patterns
  during riot events.
\newblock {\em Information Processing \& Management}, 57(1):102126, 2020.

\bibitem{metaxas2015retweets}
P.~T. Metaxas, E.~Mustafaraj, K.~Wong, L.~Zeng, M.~O'Keefe, and S.~Finn.
\newblock What do retweets indicate? results from user survey and meta-review
  of research.
\newblock In {\em {ICWSM}}, pages 658--661. {AAAI} Press, 2015.

\bibitem{nasim2013commenting}
M.~Nasim, M.~U. Ilyas, A.~Rextin, and N.~Nasim.
\newblock On commenting behavior of {F}acebook users.
\newblock In {\em {HT}}, pages 179--183. {ACM}, 2013.

\bibitem{nasim2018real}
M.~Nasim, A.~Nguyen, N.~Lothian, R.~Cope, and L.~Mitchell.
\newblock Real-time detection of content polluters in partially observable
  twitter networks.
\newblock In {\em {WWW} (Companion Volume)}, pages 1331--1339. {ACM}, 2018.

\bibitem{rizoiu2018debatenight}
M.-A. Rizoiu, T.~Graham, R.~Zhang, Y.~Zhang, R.~Ackland, and L.~Xie.
\newblock {\#}{D}ebate{N}ight: The role and influence of socialbots on
  {T}witter during the 1st 2016 {U.S. P}residential debate.
\newblock In {\em {ICWSM}}, pages 300--309. {AAAI} Press, 2018.

\bibitem{Shao_2016}
C.~Shao, G.~L. Ciampaglia, A.~Flammini, and F.~Menczer.
\newblock Hoaxy: {A} platform for tracking online misinformation.
\newblock In {\em {WWW} (Companion Volume)}, pages 745--750. {ACM}, 2016.

\bibitem{StarbirdWilson2020}
K.~Starbird and T.~Wilson.
\newblock {Cross-Platform Disinformation Campaigns: Lessons Learned and Next
  Steps}.
\newblock {\em Harvard Kennedy School Misinformation Review}, Jan. 2020.

\bibitem{Stilgherrian2020zdnet}
Stilgherrian.
\newblock {Twitter bots and trolls promote conspiracy theories about Australian
  bushfires | ZDNet}.
\newblock
  https://www.zdnet.com/article/twitter-bots-and-trolls-promote-conspiracy-theories-about-australian-bushfires/,
  Jan. 2020.
\newblock (Accessed on 2020-01-28).

\bibitem{tuke2020pachinko}
J.~Tuke, A.~Nguyen, M.~Nasim, D.~Mellor, A.~Wickramasinghe, N.~Bean, and
  L.~Mitchell.
\newblock Pachinko prediction: A bayesian method for event prediction from
  social media data.
\newblock {\em Information Processing \& Management}, 57(2):102147, 2020.

\bibitem{WeberNMF2020reliability}
D.~Weber, M.~Nasim, L.~Mitchell, and L.~Falzon.
\newblock A method to evaluate the reliability of social media data for social
  network analysis.
\newblock In {\em {ASONAM}}. {ACM}, 2020.
\newblock Accepted.

\end{thebibliography}

\end{document}